\documentclass[aps,prb,twocolumn,floatfix,superscriptaddress]{revtex4-2}

\usepackage{amssymb}
\usepackage{amsmath}
\usepackage{graphicx}

\begin{document}

\title{Determination of the magnetic penetration depth with measurements of the vortex-penetration field for type-II superconductors }

\author{G.~P.~Mikitik}
\affiliation{B.~Verkin Institute for Low Temperature Physics \&
Engineering of the National Academy of Sciences of Ukraine, Kharkiv 61103, Ukraine}

\author{Yu.~V.~Sharlai}
\affiliation{B.~Verkin Institute for Low Temperature Physics \&
Engineering of the National Academy of Sciences of Ukraine, Kharkiv 61103, Ukraine}
\affiliation{Institute of Low Temperature and Structure Research, Polish Academy of  Sciences, 50-422 Wroc{\l}aw, Poland}

\begin{abstract}
Using a known distribution of the Meissner currents over the surface of an infinitely long superconducting slab with a rectangular cross section, we find an applied magnetic field at which vortices begin to penetrate into the superconductor. This vortex-penetration field is determined by an interplay of the  geometrical and Bean-Livingston barriers. The obtained results enable one to find the lower critical field and the London penetration depth from measurements of the magnetic induction on the surface of the superconducting slab, using, e.g., the micro-Hall probes.

\end{abstract}

\maketitle

\section{Introduction}

A temperature dependence of the  magnetic penetration depth $\lambda$ can shed light on the pairing state of electrons in superconductors \cite{hardy,proz06}. In particular, using the measured dependence of $\lambda$ on the temperature $T$, important information on this state were obtained for high-$T_c$  YBa$_2$Cu$_3$O$_{6.95}$ \cite{hardy,carrington} as well as for the Fe-based \cite{proz11,putzke14,shen} and  heavy-fermion
\cite{yamashita,takenaka,pang,ishihara} superconductors. Below we will discuss one of the methods used for the determination of $\lambda$ (or of the lower critical field $H_{c1}\propto \lambda^{-2}$), viz., the measurements of the magnetic field $H_p$ at which the vortices begin to penetrate into a type-II superconductor \cite{putzke14,shen,yamashita,takenaka,Lya04,pribulova,okazaki,klein,prib,juraszek20}.
In this method, it is important to correctly take into account the shape of the superconducting sample, which in most experiments is a rectangular parallelepiped. Strictly speaking, the well-known concept of the demagnetization factor is not applicable to such platelet-shaped superconductors since the  geometrical barrier \cite{ZL} appears  in samples different from ellipsoids, and the uniform penetration of the vortices into the superconductor  no longer  occurs. Although the so-called effective demagnetization factor $N$ of a superconductor in the Meissner state is frequently discussed in the literature \cite{br2,br3,proz00,pardo,proz18}, but it does not determine the penetration field. This $N$ is defined by the condition that $-VH_a/(1-N)$ is equal to  the total magnetic moment of a sample of the volume $V$ in the applied magnetic field $H_a<H_p$.

The vortex-penetration field $H_p$ for a thin strip was estimated in Ref.~\cite{ZL}, and it was shown that due to the geometrical barrier, this $H_p$ essentially exceeds the field $(1-N)H_{c1}$, at which the vortex penetration into an ellipsoid-shaped sample occurs. Based on numerical calculations of the electrodynamics of vortices in infinite slabs with rectangular cross sections,  E.H. Brandt \cite{br2}  found a formula for $H_p$ that approximately describes this field for an {\it isotropic} superconductor in a wide interval of the aspect ratios of the slabs. On the other hand, using the method of conformal mappings in the magnetostatics \cite{LL}, a distribution of currents over the surface of the infinitely long slab with a rectangular cross section was derived in the case when the slab is  in the Meissner state \cite{Meissner}. This strict result was obtained under the only assumption that $\lambda$ is much less than the width $2w$ and the thickness $d$ of the sample. Using this result, a formula for the penetration field $H_p$ was derived in the case of thin strips with $d\ll 2w$, and an interplay between the Bean-Livingston and geometrical barriers was also analyzed \cite{jetp13}. Since the result for the currents \cite{Meissner} is valid for any $d/2w$, in this paper,  we present formulas  for $H_p$ in the case of an arbitrary aspect ratio $d/2w$ of the sample, taking into account an anisotropy of the superconductor as well. The vortex pinning is assumed to be negligible (otherwise, the penetration field depends also on the pinning, and  $\lambda$ cannot be accurately found).

The paper is structured as follows: In Sec.~\ref{II} we present
the distribution of the currents in the infinitely long slab of
the rectangular cross section when the slab is  in the Meissner state.  Using this distribution, in Sec.~\ref{III} the Bean-Livingston and geometrical barrier are analyzed for anisotropic slab of the arbitrary thickness, and a simple algorithm for the calculation of $H_p$ is proposed. Within this approach,  the dependences of penetration field on the aspect ratio of the slab  and on the anisotropy of the superconductor are considered in Sec.~\ref{IV}. In Sec.~\ref{V}, we discuss how $\lambda$ or $H_{c1}$ can be extracted from various experimental data on the vortex penetration into the superconducting slab.
The obtained results are briefly summarized in Conclusions, whereas
the Appendices contain some mathematical details of the calculations.

\section{Slab in the Meissner state} \label{II}

\begin{figure}[t] 
 \centering  \vspace{+9 pt}
\includegraphics[bburx=720,bbury=480,scale=.41]{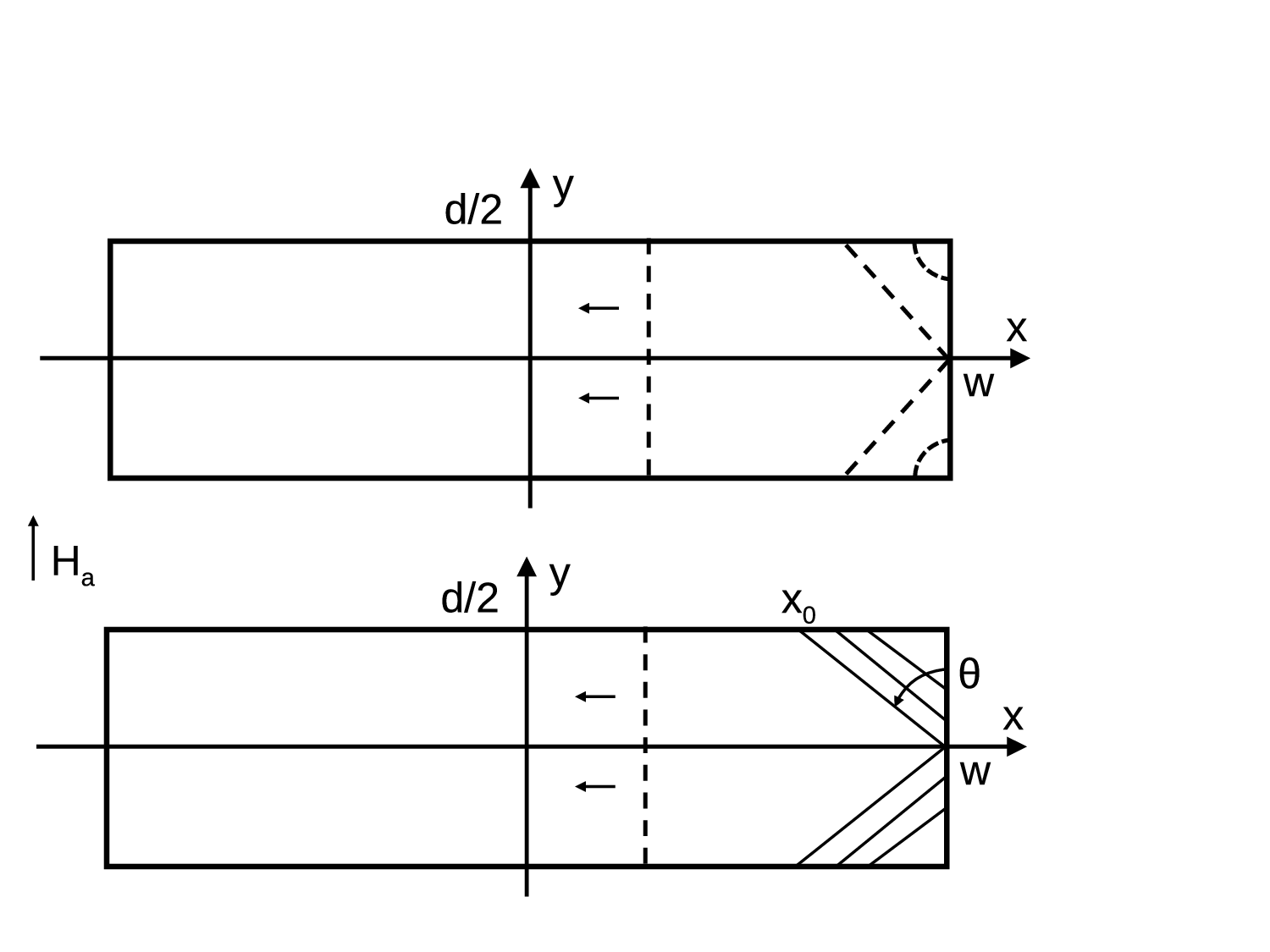}
\caption{\label{fig1} Two scenarios of the vortex penetration into the infinitely long superconducting slab of the rectangular cross section.  The thickness of the slab is $d$, and its width is $2w$. Top: $p>p_c$, the Bean-Livingston barrier prevails over the geometrical one. Bottom: $p<p_c$, the penetration of vortices is mainly determined by the geometrical barrier. The parameter $p$ is defined by Eq.~(\ref{22}), its critical value $p_c$ depends on the aspect ratio $d/2w$ of the slab and the anisotropy of the superconductor $\varepsilon$, Eq.~(\ref{23}).  The dashed lines schematically show mobile vortices in the slab, whereas the solid lines inside the slab designate the immobile vortices that are in the equilibrium. These inclined vortices form the flux-line domes on the upper (lower) plane of the slab and on its right (left) lateral surface.
 } \end{figure}

Consider a superconducting slab of a rectangular cross section of width $2w$ ($-w\le x\le w$) and thickness $d$ ($-d/2 \le y \le d/2$), which infinitely extends in the $z$ direction (Fig.~\ref{fig1}). The slab is subjected to a perpendicular applied magnetic field ${\bf H}_a=(0,H_a,0)$. It is always assumed below that $d, w \gg \lambda$.

When a superconductor is in the Meissner state, the total magnetic field ${\bf H}$ at its surface is tangential to this surface. This  ${\bf H}(x,y)$ outside the sample, and hence the Meissner sheet currents $J_M=J_z$ flowing  in the surface layer of the thickness  $\sim \lambda$, can be found by a conformal mapping \cite{LL}.  For the slab, the Meissner currents were obtained in Ref.~\cite{Meissner} (the appropriate mapping was detailed in the Supplemental Material to Ref.~\cite{prb21}). We now present the results of Ref.~\cite{Meissner} that will be necessary in subsequent calculations.

Due to the symmetry of the slab,
it is sufficient to deal with a quarter of its surface ($x \ge 0$, $y\ge 0$) and to parameterize the surface with the single variable $u$ changing from $0$ to $1/\sqrt{1- m}$ \cite{prb21}. Here $m$ is a constant parameter, $0\le m \le 1$, the value of which is determined by the aspect ratio of the slab, $d/2w$. In particular, the upper surface of the slab ($0\le x\le w$, $y=d/2$) is  parameterized as follows ($0\le u \le 1$):
\begin{eqnarray} \label{1}
\frac{x}{w}=\frac{f(u,1-m)}{f(1,1-m)}
\end{eqnarray}
where
\begin{eqnarray} \label{2}
f(u,m)&\equiv& m\int_0^{u}\!\!\frac{\sqrt{1-v^2}}{\sqrt{1-mv^2}}\,dv. \end{eqnarray}
The points $u= 1$ correspond to the upper corner of the slab, $(w,d/2)$. The constant parameter $m$ is found from the equation:
\begin{eqnarray} \label{3}
\frac{d}{2w}=\frac{f(1,m)}{f(1,1-m)}.
\end{eqnarray}
At $d\ll w$,  relation (\ref{3}) gives
\begin{eqnarray} \label{4}
m\approx \frac{2d}{\pi w}.
\end{eqnarray}
The upper part of the lateral surface, ($x=w$, $0\le y\le d/2$), have the following parametric representation ($1\le u \le 1/\sqrt{1-m}$):
\begin{eqnarray} \label{5}
\frac{2y}{d}=\frac{f(s(u),m)}{f(1,m)},
\end{eqnarray}
where
\begin{eqnarray*}
s(u)=\sqrt{\frac{1-(1-m)u^2}{m}}.
\end{eqnarray*}
The value $u=1/\sqrt{1-m}$ corresponds to the
equatorial point $(w,0)$ of the slab.

The Meissner sheet currents on the upper and lateral
surfaces of the slab (i.e., in the whole interval $0\le u \le 1/\sqrt{1-m}$) are described by the unified formula:
\begin{eqnarray} \label{6}
J_M(u)=\frac{uH_a}{\sqrt{|1-u^2|}}.
\end{eqnarray}
Formulas (\ref{1})--(\ref{6}) provide the quantitative description (in the parametric form) of the surface Meissner currents in the slab, including its edge regions. Note that the above formulas describe both the case of a thin strip in the perpendicular magnetic field ($d\ll 2w$, or equivalently, $m\ll 1$) and the case of a plate in the magnetic field parallel to its surface ($d\gg 2w$, or $1-m\ll 1$).

In the limit $|1-u^2|\ll m$, i.e., at $l\equiv w-x\ll d, w$, or at $l\equiv (d/2)-y\ll d, w$, the surface current diverges like $l^{-1/3}$ near the corners of the slab \cite{Meissner,jetp13}. In this limiting case formulas (\ref{1})-(\ref{3}) and (\ref{6}) lead to the expression
\begin{eqnarray} \label{7}
J_M\!\approx\! H_a\!\left(\!\frac{(1-m)d}{6\sqrt{m}f(1,m)\,l}\! \right)^{\!1/3},~~
\end{eqnarray}
which is valid for the slab of an arbitrary thickness.  For the thin strips, expression (\ref{7}) is further simplified since  $f(1,m)\approx \pi m/4$ at $m\ll 1$. The divergence of the current in Eq.~(\ref{7}) should be cut off at $l\lesssim \lambda$, and the current density $j$ throughout the corner region ($w-\lambda \le x \le w$, $(d/2)-\lambda \le y\le d/2$) is approximately constant, $j_{\rm crn}(x,y)\sim J_M(x=w-\lambda)/\lambda$,
\begin{eqnarray} \label{8}
j_{\rm crn}\!\sim\!\frac{H_a}{\lambda }\!\left(\!\frac{(1-m)d}{6\sqrt{m}f(1,m)\,\lambda}\! \right)^{\!1/3}.~~
\end{eqnarray}
In the case of the thin strip, formula (\ref{8}) reduces to the appropriate expression of Ref.~\cite{jetp13}.

\section{Bean-Livingston and geometrical barriers in the slab}\label{III}

\subsection{Bean-Livingston barrier} \label{BL}

Since the Meissner currents are maximum at the corners of the sample, it is favorable for a vortex to penetrate into the strip through these points. A small circular vortex arc appearing in one of the corners  overcomes the Bean-Livingston barrier and begins to expand when the current density in the corners reaches the value $j_0$ \cite{jetp13},
 \begin{equation}\label{9}
j_0\approx \frac{0.92H_{c1}\kappa}{\lambda \ln\kappa},
 \end{equation}
where $H_{c1}=\Phi_0\ln\kappa/(4\pi\mu_0\lambda^2)$ is the lower critical field, $\Phi_0$ is the flux quantum, and $\kappa$ is the Ginzburg-Landau parameter. This $j_0$ is of the order of the depairing current density $j_{dp}$ \cite{bl}, whereas $j_0\lambda$, the local surface field near the corner, reaches the value of the thermodynamic critical field $H_c$ in the agreement with the results of Refs.~\cite{dezhen,samokh2,galaiko,genen}. Equating this $j_0$ with the current density $j_{\rm crn}$ defined by Eqs.~(\ref{8}), we find the order of magnitude of the applied field  at which the Bean-Livingston barrier disappears for a vortex penetrating through the corner of the sample,
 \begin{eqnarray}\label{10}
H_p^{BL}
&=&\frac{0.92\kappa H_{c1}}{\ln\kappa}\!\left(\!\frac{6\sqrt{m}f(1,m) \lambda}{(1-m)d}\! \right)^{\!1/3}\!\!\!\!\!\!\! \nonumber \\
&=&\frac{0.92\kappa H_{c1}}{\ln\kappa}\!\left(\!\frac{3\sqrt{m}f(1,1-m) \lambda}{(1-m)w}\! \right)^{\!1/3}\!\!\!\!\!\!\!.
 \end{eqnarray}
At $d\ll w$, formula (\ref{10}) reduces to the appropriate expression of Ref.~\cite{jetp13}.

If the superconductor is anisotropic, we will assume that the axis of the anisotropy coincides with the $y$ axis. This means that $\lambda_y$ differs from  the London penetration depth $\lambda_x=\lambda_z\equiv \lambda$ characteristic of the currents in the plane perpendicular to the axis. The anisotropy parameter $\varepsilon$ is defined as follows: $\varepsilon\equiv \lambda/\lambda_y$ \cite{bl}. (We  imply below that $\varepsilon<1$.) Since the field $H_p^{BL}$ is determined by the current density $j_0\sim j_{dp}\sim H_c/\lambda$  flowing along the $z$ axis,  this field is expected to be practically independent of $\varepsilon$.

\subsection{Geometrical barrier}
\label{GB}

Consider now the vortex-entry condition  caused by the
geometric barrier in the slab, assuming that a vortex has already overcome the Bean-Livingston barrier in the corner of the slab.  In this case a penetrating vortex can move towards the center of the sample only when its two inclined rectilinear segments meet at the right equatorial point ($x=w$, $y=0$), see Fig.~\ref{fig1}. Consider the vortex segment  which ends at the point $x_0$ of the upper plane of the slab and at the point $y_0$ of its lateral surface. The balance between the line tension of the vortex and the forces generated by the surface currents leads to the following equations in $x_0$, $y_0$
\cite{jetp13}:
\begin{eqnarray} \label{11}
\Phi_0J(x_0,d/2)&=&\frac{\delta E_l}{\delta x_0},  \\
\Phi_0J(w,y_0)&=&\frac{\delta E_l}{\delta y_0},  \label{12}
\end{eqnarray}
where $\Phi_0$ is the flux quantum, the sheet currents $J(x,y)$ are determined by formulas (\ref{1})--(\ref{6}), $E_l=Le_l$ is the line energy of a vortex segment of the length $L$. In an anisotropic superconductor, one has \cite{bl}:
  \begin{eqnarray} \label{13}
  e_l=\frac{\Phi_0^2\epsilon(\theta)}{4\pi\mu_0\lambda^2} \ln\left(\frac{\kappa}{\epsilon(\theta)}\right),
   \end{eqnarray}
where $\epsilon(\theta)= \sqrt{\cos^2\theta+ \varepsilon^2\sin^2\theta}$, $\varepsilon$ is the parameter of the anisotropy ($\varepsilon\le 1$), and $\theta <\pi/2$ is the tilt angle of the vortex relative to the $y$ axis, the axis of the anisotropy. There is also a geometrical relationship between $x_0$, $y_0$, and $\theta$, which is evident from Fig.~\ref{fig1}:
\begin{eqnarray} \label{14}
w-x_0=(\frac{d}{2}-y_0)\tan\theta.
\end{eqnarray}
Equations (\ref{11})--(\ref{14}) completely determine the three quantities $x_0$, $y_0$ and $\theta$, and these equations in the explicit form are presented in Appendix \ref{A}. If we set $y_0=0$, the equations give $x_0$, the appropriate angle $\theta_0$, and the penetration field $H_p^{GB}$ caused by  the geometrical barrier.

Below we will neglect the angular dependence of the logarithmic factor in  formula  (\ref{13}) for $e_l$ (i.e., we set $\epsilon(\theta)=1$ in the logarithmic factor). Under this assumption and at $y_0=0$, the explicit form of Eqs.~(\ref{11}) and (\ref{12})  looks as follows:
\begin{eqnarray}\label{15}
 \frac{\varepsilon^2\sin\theta_0}{\epsilon(\theta_0)}\!&= &\!\!\frac{H_a}{H_{c1}}
 \frac{u_0}{\sqrt{1\!-\!u_0^2}},~~  \\
 \frac{\cos\theta_0}{\epsilon(\theta_0)}\!&=&\!\!\frac{H_a}{H_{c1}\sqrt{m}}, \label{16}
  \end{eqnarray}
where the parameter $u_0$ corresponds to the point $x_0$ according to Eq.~(\ref{1}), and $H_{c1}=e_l(\theta=0)/\Phi_0$ is the lower critical field  of the superconducting material for the magnetic field parallel to the $y$ axis. With the use of formula (\ref{1}), the geometrical relationship  (\ref{14}) at $y_0=0$ takes on the form:
\begin{eqnarray}\label{17}
\frac{d}{2w}\tan\theta_0=\frac{f(1,1-m)-f(u_0,1-m)}{f(1,1-m)}.
 \end{eqnarray}

To solve Eqs.~(\ref{15})--(\ref{17}), we note that the ratio of  formulas (\ref{15}) and (\ref{16}) enables us to express $\tan\theta_0$ in terms of $u_0$. Inserting this expression into formula (\ref{17}) and using relation (\ref{3}), we arrive at the equation determining $u_0$  as a function of the parameters  $m$ and $\varepsilon$,
\begin{eqnarray}\label{18}
\frac{\sqrt{m}u_0}{\varepsilon^2\sqrt{1-u_0^2}} =\frac{f(1,1-m)-f(u_0,1-m)}{f(1,m)}.
 \end{eqnarray}
This simple equation [see formula (\ref{2})]  has a unique solution since its left hand side increases with increasing $u_0$, while the right hand side decreases. Knowing $u_0$, we find $\theta_0$ from formula (\ref{17}), whereas Eq.~(\ref{16}) gives the penetration field caused by the geometrical barrier,
 \begin{eqnarray}\label{19}
  H_p^{GB}=H_{c1}\sqrt{m}\,\frac{\cos\theta_0}{\epsilon(\theta_0)} =\frac{H_{c1}\sqrt{m}}{\sqrt{1+\varepsilon^2\tan^2\theta_0}}.
  \end{eqnarray}

Consider the two special cases. In the case of a thick anisotropic slab, for which $d/2w\gtrsim 1$ (i.e., $m\sim 1/2$) and $\varepsilon^2\ll 1$, we assume that the parameter $u_0\ll 1$. Then, we find  $\tan\theta_0\approx 2w/d$ from Eq.~(\ref{17}), whereas the ratio of Eqs.~(\ref{15}), (\ref{16}) gives  $u_0\approx \varepsilon^2\tan\theta_0/\sqrt{m}$. Thus, our assumption, $u_0\ll 1$, is really fulfilled if  $\varepsilon^2\ll 1$. In this case, formula (\ref{19}) is practically independent of $\varepsilon$ and reduces to
 \begin{eqnarray}\label{20}
 H_p^{GB}\approx H_{c1}\sqrt{m}.
 \end{eqnarray}

In the case of a thin isotropic strip when $\varepsilon=1$ and $m\ll 1$, the appropriate angle $\theta_0$ is almost independent of $m$ ($\theta_0\approx 36.5^{\circ}$, $\cos\theta_0\approx 0.8$) \cite{jetp13}, and
 \begin{eqnarray}\label{21}
 H_p^{GB}\approx 0.8\sqrt{m}H_{c1}\approx 0.9\sqrt{\frac{d}{2w}}H_{c1},
 \end{eqnarray}
where we have used formula (\ref{4}).

\subsection{Interplay of the barriers. Two scenarios of the vortex penetration} \label{IIIc}

\begin{figure}[t] 
 \centering  \vspace{+9 pt}
\includegraphics[scale=.9]{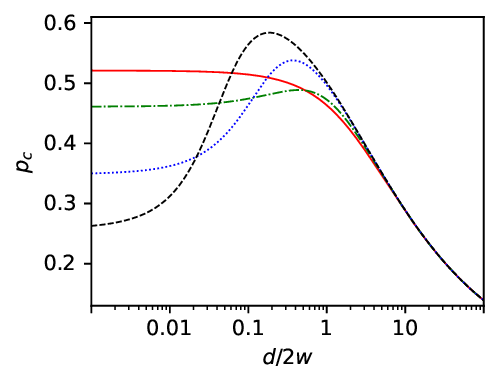}
 \caption{\label{fig2} Dependence of the critical value $p_c$, Eq.~(\ref{23}), of the parameter $p$ defined by Eq.~(\ref{22}) on the aspect ratio $d/2w$ of the superconducting slab of the thickness $d$ and of the width $2w$ for various values of the anisotropy parameter $\varepsilon$: $\varepsilon=1$ (red solid line), $\varepsilon=0.5$ (green dot-dash line), $\varepsilon=0.15$ (blue dotted line), and $\varepsilon=0.05$ (black dashed line).
 } \end{figure}

A comparison of formulas (\ref{10}) and (\ref{19}) shows that the ratio of these penetration fields  can be written as $p/p_c$ where the parameter $p$ is defined as follows:
\begin{equation}\label{22}
p\equiv \frac{\kappa}{\ln\kappa }\left (\frac{\lambda}{d}\right
)^{1/3},
\end{equation}
and $p_c$ is its  critical value which generally depends on $\varepsilon$ and $m$,
\begin{equation}\label{23}
p_c\equiv  \frac{\cos\theta_0}{0.92\epsilon(\theta_0)}\left (\frac{m(1-m)}{6f(1,m)}\right
)^{1/3}.
\end{equation}
Although $H_p^{BL}$, Eq.~(\ref{10}), and hence $p_c$ are determined up to a numerical factor of the order of unity, formula (\ref{23}) reveals the dependences of $p_c$ on $d/2w$ and $\varepsilon$.
For thin isotropic strips ($m\ll 1$, $\varepsilon=1$), this $p_c$ is practically independent of $m$ ($p_c\approx 0.52$) \cite{jetp13}. For anisotropic slabs with $d/2w\lesssim 1$, $p_c$ depends on the aspect ratio $d/2w$ and $\varepsilon$, but these dependences are relatively  weak, and the parameter $p_c$ remains of order of $0.5$, Fig.~\ref{fig2}. At $d/2w\gg 1$, $p_c$ decreases and is described by the universal function $p_c(d/2w)$ independent of $\varepsilon$,
 \[
 p_c(d/2w)\approx \frac{1}{0.92}\left(\frac{16w}{3\pi^2d}\right)^{1/3}\approx 0.7\left(\frac{2w}{d}\right)^{1/3}.
 \]

Since the parameter $p$ can be greater or less than its critical value $p_c$, two scenarios of the vortex penetration into the sample are possible \cite{jetp13}. If $p>p_c$, one has $H_p^{BL}> H_p^{GB}$, and the true penetration field $H_p$ coincides with $H_p^{BL}$, Eq.~(\ref{10}).
In this case, small vortex segments appearing at the corners of the
strip at $H_a= H_p^{BL}$ immediately expand, merge at the equatorial point ($x=w$, $y=0$), and the created vortex moves towards the center of the sample, Fig.~\ref{fig1}. If the parameter $p$ is less than the critical value $p_c$, one has $H_p^{BL}< H_p^{GB}$, and the vortex penetration is a two-stage process. The current density in the vicinity of the corners reaches the depairing value at $H_a=H_p^{BL}$. At this field a penetrating vortex enters the sample through the corner, but it cannot reach the equatorial point,
and so this vortex line will ``hang'' between the corner and the equatorial point $(x=w, y=0)$. With increasing $H_a$, two domes filled by these inclined vortex  lines will expand in the lateral surface of the strip. The penetration field $H_p$ is determined by the condition that the boundaries of these domes meet at the equatorial point, and this field  $H_p$ can be estimated by Eq.~(\ref{19}). However, formula (\ref{19}) has been derived, considering a single inclined vortex. Since the vortices ending on the lateral surface of the slab modify the current distribution in the sample, the $H_p$ has to be calculated self-consistently, taking into account the currents generated by the domes of the inclined vortices. Nevertheless, as was shown in Ref.~\cite{jetp13}, for the thin isotropic strip, the maximal decrease of $H_p$ associated with these domes does not exceed $20\%$ as compared to $H_p$ given by Eq.~(\ref{21}) [at $p\ll p_c$, the numerical coefficients $0.8$ and $0.9$ in Eq.~(\ref{21}) are replaced by $0.63$ and $0.71$, respectively]. Below, we  will neglect the currents generated by the domes since on the one hand, this neglect essentially simplifies the problem, and on the other hand, this approach provides a sufficiently accurate calculation of $H_p$.

\section{Penetration field} \label{IV}

\begin{figure}[t] 
 \centering  \vspace{+9 pt}
\includegraphics[scale=.9]{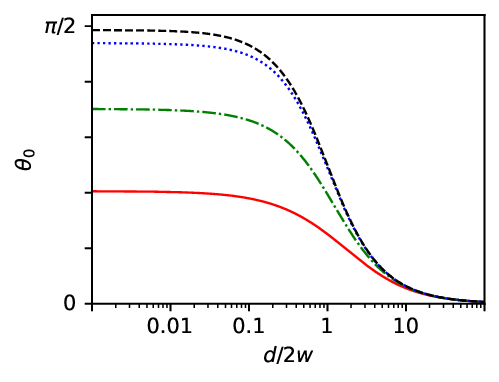}
 \caption{\label{fig3} Dependence of the angle $\theta_0$ of the inclined vortex on the aspect ratio $d/2w$ of the superconducting slab for various values of the anisotropy parameters $\varepsilon=1$, $0.5$, $0.15$, $0.05$. The correspondence of $\varepsilon$ to the color and style of the appropriate line is the same as in Fig.~\ref{fig2}. The curves are calculated, solving Eqs.~(\ref{15})--(\ref{19}) in the neglect of the angular dependence of the logarithmic factor in Eq.~(\ref{13}).
 } \end{figure}   

\begin{figure}[t] 
 \centering  \vspace{+9 pt}
\includegraphics[scale=.9]{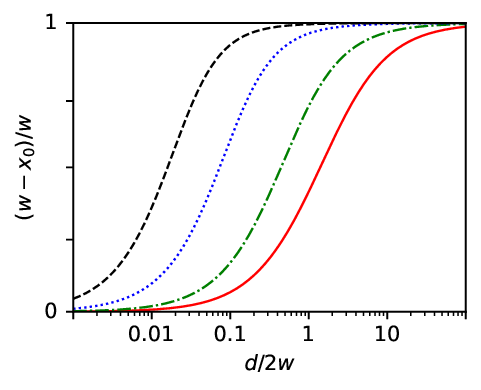}
 \caption{\label{fig4}  The position $x_0$ of the  inclined vortex on the upper plane of the slab versus the aspect ratio $d/2w$ of the superconducting slab for various values of the anisotropy parameter $\varepsilon=1$, $0.5$, $0.15$, $0.05$. The correspondence between the values of $\varepsilon$ and the lines is the same as in Figs.~\ref{fig2} and \ref{fig3}.  The curves are calculated, solving Eqs.~(\ref{15})--(\ref{19}) in the neglect of the angular dependence of the logarithmic factor in Eq.~(\ref{13}).
 } \end{figure}   

Using the equations of the preceding section, consider dependences of the vortex penetration field on the aspect ratio of the slab and on the anisotropy parameter $\varepsilon$. If the parameter $p$ exceeds its critical value $p_c$, the penetration field is estimated by formula (\ref{10}), which is independent of $\varepsilon$. To understand the dependence of this field on the aspect ratio of the slab, let us assume that its width $2w$ is constant, whereas the thickness $d$ increases. Then, formula (\ref{10}) shows that $H_p^{BL}$  is proportional to $d^{1/6}$ at small $d/2w$ and reaches its maximal value
 \begin{equation}\label{24}
 H_{p,max}^{BL}= 0.92\frac{\kappa H_{c1}}{\ln\kappa}\left(\frac{3\pi\lambda}{4w}\right)^{1/3}
  \end{equation}
in the limit $d\to \infty$. Note that this limiting value is noticeably  less than the Bean-Livingston penetration field $H_c=\sqrt{2}H_{c1}\kappa/\ln\kappa$ characteristic of flat surfaces \cite{dezhen},  since the vortices now penetrate into the samples through its corners. However, if this maximal value is less than $H_{c1}$, the curve $H_p^{BL}(d/2w)$  necessarily crosses the curve $H_p^{GB}(d/2w)$ at some $d_{\rm cr}$ (see figures below), and one has $p=p_c$ at the crossing point. When $d>d_{\rm cr}$, $H_p^{GB}(d/2w)$ becomes larger than $H_p^{BL}(d/2w)$, and the penetration field is determined by the geometrical barrier.

Consider now the situation when $p<p_c$, and the penetration field is determined by the geometrical barrier, described by  Eqs.~(\ref{17})--(\ref{19}). In Figs.~\ref{fig3} and \ref{fig4}, we show the dependences of $\theta_0$ and $(w-x_0)/w$ on the aspect ratio $d/2w$ of the slab for various values of the anisotropy parameter $\varepsilon$. Note that the angle $\theta_0$ remains practically constant  for the strips with $d/2w\lesssim 0.1$. In the isotropic case, $\varepsilon=1$, the quantity $(w-x_0)/w$, defining the position of the inclined vortex on the upper plane of the slab, becomes comparable with unity only at $d/2w> 1$. However, the less $\varepsilon$, the greater $(w-x_0)/w$, and $(w-x_0)/w\sim 1$ even for thin strips if $\varepsilon$ is small.

\begin{figure}[t] 
 \centering  \vspace{+9 pt}
\includegraphics[scale=.9]{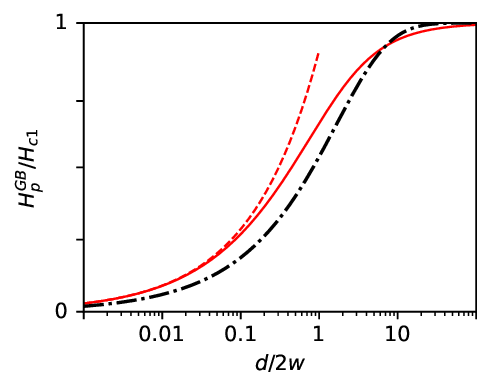}
 \caption{\label{fig5}  Dependence of the penetration field $H_p^{GB}$, caused by the geometrical barrier, on the aspect ratio $d/2w$ of the  isotropic superconducting slab ($\varepsilon=1$). The red curve is obtained, solving Eqs.~(\ref{15})--(\ref{19}) in the  neglect of the angular dependence of the logarithmic factor in Eq.~(\ref{13}). The black dot-and-dash line corresponds to formula (\ref{25}), whereas the red dashed line is described by Eq.~(\ref{21}).
 } \end{figure}

The dependence  $H_p^{GB}/H_{c1}$ on the aspect ratio $d/2w$ of the isotropic slab ($\varepsilon=1$) is presented in Fig.~\ref{fig5}. For comparison, in Fig.~\ref{fig5} we also show this dependence obtained from numerical calculations of the electrodynamics of vortices \cite{br2},
 \begin{eqnarray}\label{25}
 \frac{H_p^{GB}}{H_{c1}}\approx \tanh\left(\sqrt{0.36\frac{d}{2w}}\right).
 \end{eqnarray}
The dashed line in this figure corresponds to formula (\ref{21}) obtained for thin strips ($d\ll w$) \cite{jetp13}. It is seen that this formula well describes the function $H_p^{GB}(d/2w)$ at $d/2w\lesssim 0.1$. Interestingly, Brandt's formula (\ref{25}) in the case of the thin strips gives the expression,
 \[
 \frac{H_p^{GB}}{H_{c1}}\approx 0.6\sqrt{\frac{d}{2w}},
  \]
that differs from Eq.~(\ref{21}) only by the numerical factor of the order of unity. The difference between these thin-strip results is further reduced if we take into account that the vortex domes of the inclined vortices  lead to the decrease of the coefficient before $\sqrt{d/2w}$ in Eq.~(\ref{21}) (see the end of Sec.~\ref{IIIc}). Thus, our approach based on the solution of simple equation (\ref{18}) and Brandt's formula (\ref{25}) lead to close values of $H_p$ in the isotropic case. However, our approach also describes the anisotropic case; in addition, it permits one to analyze the situations when the complete penetration of vortices does not occur (see the next section).

\begin{figure}[t] 
 \centering  \vspace{+9 pt}
\includegraphics[scale=.9]{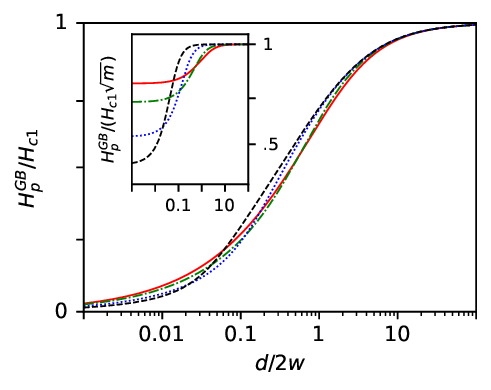}
 \caption{\label{fig6} Dependences of the penetration field $H_p^{GB}$, caused by the geometrical barrier, on the aspect ratio $d/2w$ of the  anisotropic superconducting slab  for different values of the anisotropy parameter $\varepsilon=1$, $0.5$, $0.15$, and $0.05$. The notations of the lines are the same as in Figs.~\ref{fig2}-\ref{fig4}. Inset shows $H_p^{GB}/(\sqrt{m}H_{c1})$ versus $d/2w$. This ratio $H_p^{GB}/(\sqrt{m}H_{c1})$ tends to constant values both at $d/2w\to 0$ and at $d/2w\to \infty$; see Eqs.~(\ref{19}), (\ref{20}), and Fig.~\ref{fig3}.
 } \end{figure}

In Figs.~\ref{fig6}, we show the  dependences of  $H_p^{GB}$ on $d/2w$ for various values of the anisotropy parameter $\varepsilon$. For thick slabs $d\gg 2w$, this dependence is well described by formula (\ref{20}). However, it is important that even for $d/2w\lesssim 1$, the dependences $H_p^{GB}(d/2w)$ calculated for different $\varepsilon$ are close to each other. In other words, the anisotropy of a superconducting material has a relatively small effect on $H_p^{GB}$ for not-too-thin samples. Probably, this result explains the successful applications of formula (\ref{25}) found for isotropic superconductors to the anisotropic materials \cite{pribulova,okazaki,klein}.

\section{Discussion} \label{V}

Using the micro-Hall probes, the magnetic-induction component $B_y$ is usually measured on the upper (lower) surface of the slab to detect the penetration of vortices into a type-II superconductor, and therefore, to find the penetration field $H_p$.
Apart from the Hall-probe magnetometry, such measurements can be also carried out, using ensembles of  nitrogen-vacancy centers  \cite{joshi}, magneto-optical imaging \cite{jooss}, and the nanoscale SQUID-on-tip \cite{embon}.  The formulas of this paper permit one to obtain information on $\lambda$ (or $H_{c1}$) from the $H_p$ thus measured. For definiteness, we imply below that $B_y$ is measured by the micro-Hall probes.

Consider $B_y(x)$ on the upper plane of the slab when the first scenario of the vortex penetration into the sample occurs (i.e., when $H_p^{BL}>H_p^{GB}$), Sec.~\ref{IIIc}. In this case,  a penetrating vortex arrive at the center of the slab at $H_a=H_p^{BL}$, and the nonzero $B_y$ can be detected only by the sensor placed at this point. At $H_a>H_p^{BL}$, the vortices accumulate in the center of the slab, and the vortex dome on its upper plane gradually extends. This dome is described by the formula \cite{prb21}:
 \begin{eqnarray}\label{26}
 B_y(u)=\mu_0H_a\frac{\sqrt{u_d^2-u^2}}{\sqrt{1-u^2}},
 \end{eqnarray}
where $u^2\le u_d^2$, and the boundary of the vortex dome $u_d$ is determined by the condition \cite{prb21} that the current density at $x=w$ is equal to $j_0$ defined by Eq.~(\ref{9}). This condition yields
 \begin{eqnarray}\label{27}
 u_d^2=1-\left(\frac{H_p^{BL}}{H_a}\right)^2.
 \end{eqnarray}
When $u_d\ll 1$ and $m\ll 1$, formula (\ref{1}) gives $u\approx x/w$, and  expressions (\ref{26}), (\ref{27}) reduces to the well-known result for a thin strip \cite{ZL}. However, Eqs.~(\ref{26}) and (\ref{27}) are applicable to the slab with an arbitrary aspect ratio $d/2w$, and one can easily find the field $H_a(x_s)$ at which the boundary of the dome arrives at a point $x_s$, i.e., the field at which a Hall sensor placed at this point   begins to show a nonzero signal. Note that the signal sharply grows at $H_a\gtrsim H_a(x_s)$. For example, Eqs.~(\ref{26}) and (\ref{27}) at $u=0$ give, $d(B_y/\mu_0)/dH_a= H_a/\sqrt{H_a^2-(H_p^{BL})^2}$. This derivative is large at $H_a\approx H_p^{BL}$. Thus, in the case of the first scenario, the measurement of the penetration field  permits one to find $H_p^{BL}$ determined by formula (\ref{10}), i.e., to estimate the combination of parameters,  $\kappa/\lambda^{5/3}$. However, it is necessary to keep in mind that local defects of the corners of the slab are favorable for the vortex penetration through these defects \cite{aladyshkin}, and the measured $H_p$ can be noticeably suppressed as compared to the theoretical value.

If the second scenario of the vortex penetration occurs (i.e., if  $H_p^{BL}<H_p^{GB}$), a vortex arrives at the center of the slab at $H_a=H_p^{GB}$, and for $H_a>H_p^{GB}$ the physical picture is qualitatively the  same as for the first scenario. However, formulas  (\ref{26}) and (\ref{27}), strictly speaking, should be modified. This modification is due to the currents generated by the inclined vortices that hang near the corners. Within our simplified approach, which neglects these currents, the penetration field $H_p^{GB}$ is determined by formula (\ref{19}). Thus, the measurement of this field gives $H_{c1}$ if the anisotropy parameter $\varepsilon$ is known. It is important to emphasize that, in contrast to $H_p^{BL}$, the penetration field $H_p^{GB}$ is insensitive to small defects of the surface of the slab, since the geometric barrier is determined by the length of the order of the thickness $d$ of the sample.

\begin{figure}[t] 
 \centering  \vspace{+9 pt}
\includegraphics[scale=.9]{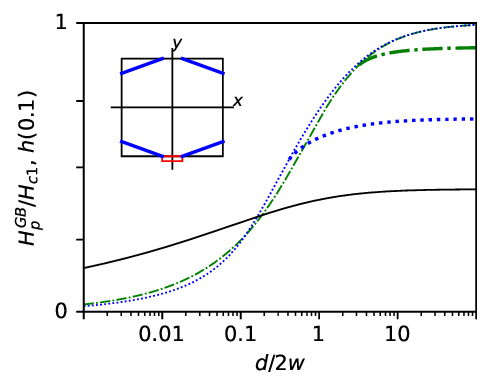}
\caption{\label{fig7} Dependences of the dimensionless magnetic field $h(x_s/w)=H_a/H_{c1}$, at which the inclined vortex arrives at the point $x_s=0.1w$, on the aspect ratio $d/2w$ of the  anisotropic superconducting slab for the two values of the anisotropy parameter $\varepsilon=0.5$ (green thick dot-and-dash line) and $\varepsilon=0.15$ (blue thick dotted line). The similar thin lines show $H_p^{GB}/H_{c1}$ versus $d/2w$, cf. Fig.~\ref{fig6}. The dependence $H_p^{BL}(d/2w)/H_{c1}$, Eq.~(\ref{10}), is shown by the black solid line for the case $H_{p,max}^{BL}=0.424H_{c1}$, Eq.~(\ref{24}). Inset: The slab and the inclined vortices shown schematically at $H_a=H_{c1}h(x_s/w)$. At this field, the vortices reach the Hall sensor (red small rectangle) that covers the lower plane of the slab from $x=-x_s$ to $x=x_s$.
 } \end{figure}

However, it is clear that the inclined vortices can generate a nonzero $B_y$ on the surface of the sample if $H_p^{BL}<H_a<H_p^{GB}$, i.e., when the second end of the inclined vortex has not yet reached the equatorial point of the slab (Fig.~\ref{fig7}, inset).  Taking into account the data of Fig.~\ref{fig4}, we conclude that this nonzero $B_y$ can be detected on the upper plane of the slab at a sufficiently far distance ($\sim w$) from the corners if $d\gtrsim w$ or $\varepsilon^2\ll 1$. Thus, in principle, it is possible to find $H_{c1}$  not only from the field $H_p^{GB}$ of the complete vortex penetration into the sample but also from  a detection of nonzero $B_y$ generated by the inclined vortices.

The dome of the inclined vortices appears on the upper plane at $H_a=H_p^{BL}$, and it  expands with increasing $H_a$, reaching its maximal size at $H_a=H_p^{GB}$. To calculate the size of the dome correctly, it is necessary to take into account the currents produced by the inclined vortices \cite{jetp13}. Within our simplified approach, in which these currents are disregarded, the boundaries of the dome can be estimated  only approximately, calculating $x_0$ with Eqs.~(\ref{a1})--(\ref{a3}) at $H_a=H_p^{BL}$ and at given $H_a<H_p^{GB}$. Note that the smaller $(H_a-H_p^{BL})/H_p^{GB}$, the better accuracy of this estimate. Thus, if the Hall sensor placed at a point $x_s$ contacts with a small dome of the inclined vortices, equations (\ref{a1})--(\ref{a3}) with the fixed  $x_0=x_s$ permit one to calculate the value of $H_a/H_{c1}\equiv h(x_s/w)$, at which this contact occurs. Comparing the calculated $h(x_s/w)$ and the measured $H_a$, for which a nonzero $B_y$ is detected by the sensor, one can estimate $H_{c1}$.

\begin{figure}[t] 
 \centering  \vspace{+9 pt}
\includegraphics[scale=.9]{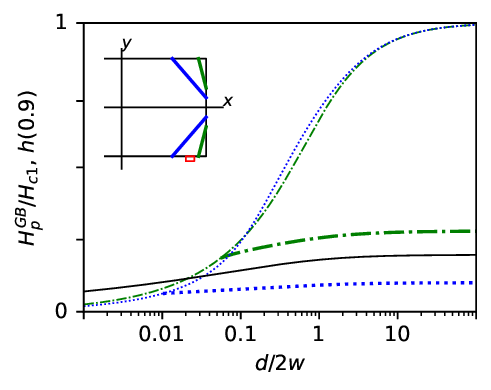}
\caption{\label{fig8} Dependences of the dimensionless magnetic field $h(x_s/w)=H_a/H_{c1}$, at which the inclined vortex arrives at the point $x_s=0.9w$, on the aspect ratio $d/2w$ of the  anisotropic superconducting slab for the two values of the anisotropy parameter $\varepsilon=0.5$ (green thick dot-and-dash line) and $\varepsilon=0.15$ (blue thick dotted line). The similar thin lines show $H_p^{GB}/H_{c1}$ versus $d/2w$, cf. Fig.~\ref{fig6}. The dependence $H_p^{BL}(d/2w)/H_{c1}$ is shown by the black solid line for the case $H_{p,max}^{BL}=0.197H_{c1}$. Inset: The inclined vortices are schematically depicted in the slab at $H_a=H_p^{BL}$.  For $\varepsilon=0.5$, the vortices (green solid lines) have not reached the micro-Hall sensor (red rectangle) yet, whereas for $\varepsilon=0.15$, the vortices (blue solid lines) have already passed through the sensor.
 } \end{figure}   

As an example, consider the case when the sensor is placed at the center of the slab, and it covers the region $-x_s\le x \le x_s=0.1w$ on the surface. In Fig.~\ref{fig7}, for different values of $d/2w$ and $\varepsilon$, we show the dimensionless magnetic fields $h(x_s/w)$, at which the end of the inclined vortex reaches the point $x_s$. The $d/2w$ dependences of these fields start on the lines $H_p^{GB}(d/2w)/H_{c1}$  since at low values of the aspect ratio, only the vertical vortex created at $H_a=H_p^{GB}$ can reach this sensor. Thus, for $\varepsilon=0.15$, Fig.~\ref{fig7} demonstrates that at $d<d_c\approx 0.4w$, the vertical vortices begin to arrive at the sensor when  $H_a=H_p^{BL}(d/2w)$. In the interval $d_c<d\lesssim 0.8w$, such vortices reach the sensor at $H_a=H_p^{GB}(d/2w)$, and for $d>0.8w$, the end of the incline vortex come to the sensor at the magnetic field $H_a$ determined by the blue thick dotted line. Note that due to the large tilt angle $\theta$, the magnitude of $B_y$ produced by the inclined vortex at the sensor is expected to be smaller than at $H_a= H_p^{GB}$ when the vertical vortices  completely penetrate into the sample.  Thus, the cases of the partial ($H_a<H_p^{GB}$) and complete ($H_a =H_p^{GB}$) vortex penetration can differ in the measured values of $dB_z/dH_a$.

Consider now the situation when the Hall sensor is near the corner of the slab. In Fig.~\ref{fig8}, we consider the case when a very small sensor is at the point $x_s=0.9w$. In this situation, for the end of the inclined vortex to be at the point $x_s$, the field $H_p^{BL}$ has to be sufficiently small, and $H_a$ should only slightly exceeds $H_p^{BL}$. Otherwise,  the vortex dome of such vortices can be between the sensor and the center of the slab. It is this case that occurs for $\varepsilon=0.15$ in Fig.~\ref{fig8}. In this case, the blue thick dotted line lies below $H_p^{BL}(d/2w)$, and the inclined vortices cannot come to the sensor. On the other hand, for $\varepsilon=0.5$, the green thick dot-and-dash line shows the dimensionless fields $h(0.9)$ at which the inclined vortex reaches the sensor.

\begin{figure}[t] 
 \centering  \vspace{+9 pt}
\includegraphics[scale=.9]{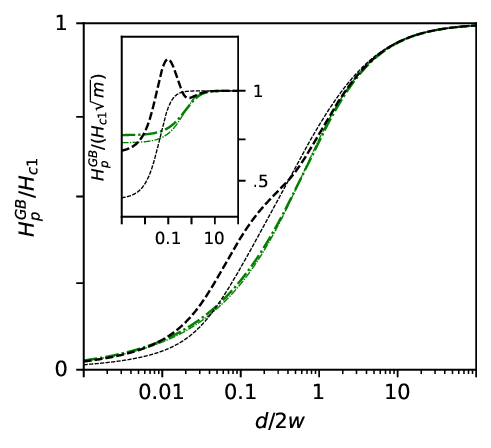}
 \caption{\label{fig9} Dependences of the penetration field $H_p^{GB}$ on  the aspect ratio $d/2w$ of the  anisotropic superconducting slab for two values of the anisotropy parameter $\varepsilon=0.5$ (green dot-and-dash lines) and $0.05$ (black dashed lines). The thick lines are calculated without taking into account the angular dependence of the logarithmic factor in formula (\ref{13}), cf. Fig.~\ref{fig6}. The similar thin lines are calculated with Eqs.~(\ref{b1}), (\ref{b2}), (\ref{17}) and $\kappa=50$. Inset shows the ratio $H_p^{GB}/(\sqrt{m}H_{c1})$ versus $d/2w$.
 } \end{figure}   

In deriving Eqs.~(\ref{17})-(\ref{19}), we neglected the angular dependence of logarithmic factor in the vortex energy $e_l$, Eq.~(\ref{13}). In general, this factor depends on the vortex core. To clarify  how the core can manifest itself in the dependence $H_p^{GB}(d/2w)$,  we derive the  appropriate equations in Appendix \ref{B}, using formula (\ref{13}) without any simplifications.
In Fig.~\ref{fig9}, the functions $H_p^{GB}(d/2w)$ obtained with and without taking into account the angular dependence of the logarithmic factor are compared. It is seen that this angular dependence clearly manifests itself only at $d/2w\lesssim 0.3$ for superconductors with strong anisotropy. However, we do not find  qualitative changes in the functions $H_p^{GB}(d/2w)$. For thick samples or for $\varepsilon\gtrsim 1/2$, the changes are small.

Throughout this paper, we have considered the infinite slab along the $z$ axis. In Appendix \ref{C}, we discuss how one can approximately take into account the finite length $L$ of real platelet-shaped samples.

\section{Conclusions}

We analyze the process of the vortex penetration into an anisotropic superconducting slab of a rectangular cross section, with the
width $2w$ and the thickness $d$ of the sample being much larger than the London penetration depth $\lambda$. The aspect ratio of the slab, $d/2w$, may have an arbitrary value. The axis of the anisotropy of the superconductor and the applied magnetic field $H_a$ are assumed to be directed along the thickness of the sample, and the flux-line pinning is neglected. It is shown that the vortex-penetration field $H_p$ coincides with the largest field from $H_p^{BL}$ and $H_p^{GB}$ determined by the Bean-Livingston and geometrical barriers, respectively, and the formulas for these $H_p^{GB}$ and $H_p^{BL}$ are derived. The samples, for which the geometrical barrier dominates over the Bean-Livingston one (i.e., when $H_p^{GB}>H_p^{BL}$), are most suitable for determining $H_{c1}$ and $\lambda$, since the Bean-Livingston barrier is sensitive to the defects in the corners of the slab.  In the case $H_p^{GB}>H_p^{BL}$, the vortex penetration is a two-stage process. In the interval $H_p^{BL}<H_a<H_p$, the inclined vortices enter the edge regions of the slab through its corners, and domes of such vortices appear on the upper (lower) planes of the slab. For this case, we  describe the simple approach for calculating the field $H_a$ at which the inclined vortex comes to a given point on the upper (lower) plane of the slab.

The obtained results permit one to estimate the values of $H_{c1}$ and $\lambda$  from measurements of the penetration field  or of the field at which the inclined vortices reach a micro-Hall probe placed on upper (lower) plane of the slab.

\begin{acknowledgments}
This research was supported in part by the Program PAS-NASU 2024.
\end{acknowledgments}

\appendix

\section{Equations for the inclined vortex }\label{A}

Consider a vortex, the ends of which are at the point $x_0$ of the upper plane of the slab and at the point $y_0$ of its lateral surface. Then, Eqs.~(\ref{11})--(\ref{14}) takes on the form [we still set $\epsilon(\theta)=1$ in the logarithmic factor of formula (\ref{13})]:
\begin{eqnarray}\label{a1}
 \frac{\varepsilon^2\sin\theta}{\epsilon(\theta)}\!&=&\!\!\frac{H_a}{H_{c1}}
 \frac{u_0}{\sqrt{1\!-\!u_0^2}},~~  \\
 \frac{\cos\theta}{\epsilon(\theta)}\!&=&\!\!\frac{H_a}{H_{c1}\sqrt{m}}
 \frac{\sqrt{1-ms_0^2}}{\sqrt{1-s_0^2}}, \label{a2}
  \end{eqnarray}
where the parameter $u_0$ corresponds to the point $x_0$,  Eq.~(\ref{1}), whereas $s_0$ corresponds to the point $y_0$, Eq.~(\ref{5}). With the use of formula (\ref{1}) and the ratio of Eqs.~(\ref{a1}) and (\ref{a2}), which gives an expression for $\tan\theta$ in terms of $u_0$ and $s_0$, the geometrical relationship  (\ref{14}) reduces to
\begin{eqnarray}\label{a3}
[f(1,1-m)&-&f(u_0,1-m)]\frac{\sqrt{1-u_0^2}\varepsilon^2}{\sqrt{m}u_0} \nonumber \\ &=&\frac{\sqrt{1-s_0^2}}{\sqrt{1-ms_0^2}}[f(1,m)-f(s_0,m)].
 \end{eqnarray}
At $s_0=0$, Eqs.~(\ref{a1})--(\ref{a3}) reduce to equations (\ref{15}), (\ref{16}), (\ref{18}). Formulas (\ref{a1})--(\ref{a3}) permit one to find any three of the four quantities: $u_0$ (i.e., $x_0$), $s_0$ (i.e., $y_0$), the angle $\theta$, and $H_a$. For example, at given $u_0$, the relation (\ref{a3}) is equation in $s_0$. Knowing $s_0$, we find the angle $\theta$ and the appropriate $H_a$ from Eqs.~(\ref{a1}) and (\ref{a2}).

\section{Penetration field $H_p^{GB}$ for a superconductor with the anisotropic vortex core}\label{B}

With the angular dependence of the logarithmic factor in  formula (\ref{13}), equations (\ref{15}), (\ref{16}) takes on the form:
\begin{eqnarray}\label{b1}
 \frac{\sin\theta_0}{\epsilon(\theta_0)}\frac{\left[\varepsilon^2 \ln(\kappa/e\epsilon(\theta_0))\!+\![\epsilon(\theta_0)]^2\right]} {\ln\kappa}\!&=&\!\!
 \frac{H_au_0}{H_{c1}\sqrt{1\!-\!u_0^2}},~~~~~~\\
 \frac{\cos\theta_0}{\epsilon(\theta_0)} \frac{\left[\ln(\kappa /e\epsilon(\theta_0))\!+\![\epsilon(\theta_0)]^2\right]}{\ln\kappa}\!&= &\!\!\frac{H_a}{H_{c1}\sqrt{m}}, \label{b2}
  \end{eqnarray}
where $\kappa$ is the Ginzburg-Landau parameter. Solving the set of  equations (\ref{b1}), (\ref{b2}), and (\ref{17}), we can find $\theta_0$, $u_0$, $H_p^{GB}$.

\section{Slab of a finite length}\label{C}

In Ref.~\cite{proz18}, the effective demagnetization factor $N$ for a platelet-shaped superconductor with dimensions $d\times 2w\times L$ was considered, and the following formula  was proposed:
\begin{equation}\label{c1}
  N(r,R)\approx \frac{4}{4+3r(1+R)},
\end{equation}
where  $r\equiv d/2w$ and $R\equiv 2w/L$ are the aspect ratios of the cross sections of this platelet. In the limiting case of the infinitely long slab ($L\to \infty$), Eq.~(\ref{c1}) yields
\begin{equation}\label{c2}
  N(r,0)\approx \frac{4}{4+3r}.
\end{equation}
Since the currents in the Meissner state of the platelet-shaped superconductor of the volume $V$ and its magnetic moment $-H_aV/(1-N)$ are proportional to each other, we can express the Meissner currents in a slab of a finite length $L$ via the currents in the infinitely long slab with the same aspect ratio $r$,
\begin{equation}\label{c3}
  \frac{J_M(r,R)}{J_M(r,0)}=\frac{1-N(r,0)}{1-N(r,R)}\equiv \frac{1}{F(r,R)}.
\end{equation}
On the other hand, the smaller the Meissner currents, the larger the  $H_p^{BL}$ and $H_p^{GB}$ determined by these currents. Therefore, to estimate effect of the finite length of the slab on the penetration fields, one can introduce the additional factor $F(r,R)$,
\begin{equation}\label{c4}
  F(r,R)=1+\frac{4R}{4+3r(1+R)},
\end{equation}
into formulas (\ref{10}) and (\ref{19}) for $H_p^{BL}$ and $H_p^{GB}$, respectively.

{}

\end{document}